\documentclass[aps,11pt,prd,showpacs,showkeys,preprintnumbers,superscriptaddress,nobibnotes,floatfix,longbibliography,notitlepage,nofootinbib]{revtex4-1}
\pdfoutput=1

\usepackage{afterpage}
\usepackage{placeins}
\usepackage{amsmath}
\usepackage{amsfonts}
\usepackage{bm}
\usepackage{amssymb}
\usepackage{mathrsfs}
\usepackage{cancel}
\usepackage{accents}
\usepackage{mciteplus,slashed}
\usepackage{amssymb,cancel,amsmath,relsize}
\usepackage{mathrsfs} 
\usepackage{dcolumn}
\usepackage{soul}
\usepackage{bm}
\usepackage[caption=false]{subfig} 
\usepackage{appendix}
\usepackage[T1]{fontenc}	
\usepackage{csvsimple}
\usepackage{hyperref}
\usepackage[capitalise]{cleveref}
\usepackage{booktabs}
\usepackage{graphicx}
\usepackage{mathrsfs}
\usepackage{floatpag}
\usepackage[utf8]{inputenc}
\usepackage[dvipsnames]{xcolor}
\hypersetup{
    colorlinks,
    linkcolor={red!50!black},
    citecolor={blue!50!black},
    urlcolor={blue!80!black}
}
\usepackage[normalem]{ulem}
\usepackage{cleveref}
\usepackage{fancybox}
\usepackage{array}
\newcolumntype{P}[1]{>{\centering\arraybackslash}p{#1}}
\newcolumntype{M}[1]{>{\centering\arraybackslash}m{#1}}
\usepackage{titlesec}

\newcommand{\be}{\begin{eqnarray*}}
	\newcommand{\ee}{\end{eqnarray*}}

\newcommand{\bee}{\begin{eqnarray}}
	\newcommand{\eee}{\end{eqnarray}}
\newcommand{\beeq}{\begin{equation}}
	\newcommand{\eeq}{\end{equation}}

\newcommand {\ignore}[1]{}

\newcommand{\bmt}{\begin{pmatrix}}
\newcommand{\emt}{\end{pmatrix}}
\newcommand{\ba}{\begin{array}{c}}
\newcommand{\ea}{\end{array}}
\newcommand{\bea}{\begin{eqnarray}}
\newcommand{\eea}{\end{eqnarray}}

\newcommand{\bi}{\begin{itemize}}
\newcommand{\ei}{\end{itemize}}

\newcommand{\baz}{\begin{array}{cc}}
\newcommand{\besub}{\begin{subequations}}
\newcommand{\eesub}{\end{subequations}}

\titleformat*{\paragraph}{\bfseries \itshape}
\titleformat*{\section}{\centering\bfseries }
\titleformat*{\subsection}{\centering\bfseries }

\graphicspath{{./figures/}}

\begin{document}
\title{\Large Freeze-In Dark Matter and Leptogenesis: a $\psi'$SM route}

\author{Adeela Afzal}
\affiliation{Bogoliubov Laboratory of Theoretical Physics, Joint Institute for Nuclear Research 141 980, Dubna, Moscow Region, Russia}

\affiliation{National Centre for Physics (NCP), Quaid-i-Azam University Campus, Shahdra Valley Road, Islamabad 44000, Pakistan}

\author{Rishav Roshan}
\affiliation{School of Physics and Astronomy, University of Southampton, Southampton SO17 1BJ, United Kingdom}


\begin{abstract}
We investigate the possibility of \emph{freeze-in} dark matter production and baryogenesis via leptogenesis in a $\psi'$SM model, which is an $E_6$ extension of the Standard Model, featuring a residual $U(1)_{\psi'}$ gauge symmetry. This symmetry arises from a linear combination of $U(1)_\chi$ and $U(1)_{\psi}$, both of which are subgroups of the $E_6$. The spontaneous breaking of $U(1)_{\psi'}$ symmetry governs the dynamics of a singlet fermion, which serves as a freeze-in dark matter candidate. The dark matter mass arises from dimension-five operators, and a discrete symmetry ensures its stability. We show that freeze-in production from scalar decay can yield the correct relic abundance for dark matter masses between a few MeV to a few hundred GeV. Simultaneously, heavy right-handed neutrinos generate light neutrino masses via the type-I seesaw and produce the observed baryon asymmetry via leptogenesis.

\end{abstract}

\maketitle


\section{Introduction}
    \label{sec.introduction}
It is well established that the total matter content of the Universe is dominated by a non-baryonic form of matter known as dark matter (DM), while visible matter accounts for only about $5\%$ of the total energy density.
Since the Standard Model (SM) fails to address these two longstanding cosmological puzzles, several beyond the SM (BSM) have been proposed in the past few decades. Among them, Weakly Interacting Massive Particles (WIMPs) \cite{Kolb:1990vq, Jungman:1995df, Bertone:2004pz, Feng:2010gw, Arcadi:2017kky, Roszkowski:2017nbc} have long been considered the leading DM candidates, while baryogenesis via leptogenesis \cite{Weinberg:1979bt, Kolb:1979qa, Fukugita:1986hr} remains the most attractive mechanism to explain the observed matter–antimatter asymmetry in the Universe. Despite being widely accepted, the WIMP paradigm has come under increasing scrutiny in recent years due to its null detection in both collider searches \cite{Kahlhoefer:2017dnp} and direct detection experiments \cite{LZ:2022lsv}. This has motivated the investigation of alternative DM candidates, such as the feebly interacting massive particle (FIMP) \cite{Hall:2009bx,Bernal:2017kxu,Bhattacharya:2019tqq,Barman:2020plp,Barman:2021tgt,Datta:2021elq,Bhattacharya:2021jli,King:2023ztb,Bhattacharya:2023kws,Barman:2024ujh,King:2024idj,Barman:2024nhr,Barman:2024tjt,Borah:2025ema}, where the dark matter abundance is produced via the freeze-in mechanism due to its feeble interactions with the thermal bath, a scenario that forms the core of our study.

It is known that sub-group $SO(10)\times U(1)_\psi$ of $E_6$ \cite{Gursey:1975ki,Achiman:1978vg,Shafi:1978gg} can be further decomposed to $SU(5)\times U(1)_\chi\times U(1)_\psi$ \cite{Lazarides:2019xai}. The two $U(1)$s can then be combined to form $U(1)_{\psi'}$ \cite{Lazarides:2019xai}. While the singlet fermions $\mathcal{N}_i$, from $SO(10)$ are charged under the $U(1)_{\psi'}$, the popular right-handed neutrinos (RHN) that resides in the matter $16-$plet of $SO(10)$ transformed as a singlet under $U(1)_{\psi'}$. This extension has been discussed in the context of the minimal supersymmetric standard model (MSSM) and is popularly known as $\psi'$MSSM \cite{Hebbar:2017fit}. Here, we consider a non-supersymmetric version of this extension and refer to it as $\psi'$SM. This breaking also gives rise to the formation of current-carrying metastable cosmic strings (CSs) as discussed in \cite{Afzal:2023kqs} that can radiate their energy in form of the stochastic gravitational wave background \cite{Roshan:2024qnv}. In this context, it is both timely and well-motivated to explore non-supersymmetric extensions, especially given the lack of experimental evidence for supersymmetric particles. Non-supersymmetric models not only offer a minimalistic and economical approach by avoiding the introduction of a large superpartner spectrum and associated fine-tuning issues, but also broaden the scope for addressing fundamental puzzles in particle physics and cosmology.

This $\psi'$SM model naturally features an attractive DM candidate in the form of lightest singlet fermions $\mathcal{N}_1$ from $SO(10)$ Additionally, the out of equilibrium decay of RHNs to the SM particle generates the matter-antimatter asymmetry of the universe through resonant leptogenesis \cite{Pilaftsis:1997jf,Pilaftsis:2003gt} simultaneously accounting for neutrino masses and mixing via type-I seesaw mechanism \cite{Minkowski:1977sc,Gell-Mann:1979vob,Mohapatra:1979ia,Schechter:1980gr,Schechter:1981cv}. 
Following the discussion of \cite{Hebbar:2017fit}, to ensure the stability of DM, we impose a discrete $Z_2$  symmetry \footnote{Such a $Z_2$ symmetry can arise as a remnant of the underlying $E_6$ gauge symmetry after spontaneous symmetry breaking (see, for example, \cite{Maji:2024pll}). Since the DM phenomenology remain unchanged regardless of the specific origin of $Z_2$, we do not discuss this aspect in detail here. } under which $\mathcal{N}_1$ is odd. Ref.~\cite{Hebbar:2017fit} treats 
$\mathcal{N}_1$ as cold DM produced via freeze-out in a supersymmetric setup, we explore its production via the freeze-in mechanism in a non-supersymmetric framework. In contrast to freeze-out, the DM remains out of thermal equilibrium due to its extremely feeble interactions.

This work thus provides a unified, non-supersymmetric $E_6$-derived framework ($\psi'$SM) where a singlet fermion freeze-in DM scenario is realized, and simultaneously accommodates neutrino masses and baryogenesis via leptogenesis. The careful analysis of the allowed parameter space, the consideration of freeze-in DM production, and the possibility of resonant leptogenesis at low reheating temperatures set this study apart from earlier works.

The paper is organized as follows. In Section \ref{sec:model}, we summarize the salient features of the model, including the field content and the symmetry-breaking pattern. The DM phenomenology is discussed in Section \ref{sec:DM}, and the baryogenesis via leptogenesis is discussed in Section \ref{sec:NM_lepto}. Our conclusions are summarized in Section \ref{sec:conclusion}.
\section{Model}
\label{sec:model}
We consider a $\psi'$SM model based on the gauge group $G_{\rm SM}\times U(1)_{\psi'}$ where $G_{\rm SM}=SU(3)_c\times SU(2)_L\times U(1)_Y$. Next we discuss, how one can obtain a $\psi'$SM model in an $E_6$ extension of the SM. Based on the fundamental representation of $E_6$, the $15$ chiral matter fields per family of the SM are embedded in $27$.
 The quantum numbers of these $27$ fields under the SM gauge group $G_\text{SM}$ are listed in Table~\ref{tab:fields}. In addition to the well-known RHN $N_i^c$, the representation includes components corresponding to $10$-plets of 
$SO(10)$, specifically ($H_{ui},\;H_{di},\;D_{i},\;D_{i}^c$) $-$ as well as $SO(10)$ singlets $\mathcal{N}_i$, where $i\;(=1,2,3)$ indices the family.
\begin{table}[h!]
\centering
\scalebox{0.9}{
\begin{tabular}{|c|c|c|c|c|c|}

\hline
\hline
{\textbf{Matter fields}} & {\textbf{Representations under $G_{\rm SM}$}} & {$2\sqrt{10}\,Q_{\chi}$} & {$2\sqrt{6}\,Q_{\psi}$} & {\textbf{$2\sqrt{10}\,Q_{\psi'}$}} & {$Z_2$} \\

{} & {} & {} & {} & {} & {}\\

\hline

{$q_i$} & {$({\bf 3, 2}, 1/6)$} & $-1$ & $1$ & $1$& +\\

\hline

{$u_i^c$} & {$({\bf \bar 3, 1},-2/3)$} & $-1$ & $1$ & $1$& +\\

\hline

{$d_i^c$} & {$({\bf \bar 3, 1},1/3)$} & $3$ & $1$ & $2$& +\\

\hline

{$l_i$} & {$({\bf 1, 2}, -1/2)$} & $3$ & $1$ & $2$& +\\

\hline

{$N_i^c$} & {$({\bf 1, 1}, 0)$} & $-5$ & $1$ & $0$& +\\

\hline

{$e_i^c$} & {$({\bf 1, 1}, 1)$} & $-1$ & $1$ & $1$& +\\

\hline

{$H_{ui}$} & {$({\bf 1, 2},1/2)$} & $2$ & $-2$ & $-2$& +\\

\hline

{$H_{di}$} & {$({\bf 1, 2},-1/2)$} & $-2$ & $-2$ & $-3$& +\\

\hline

{$D_i$} & {$({\bf 3, 1},-1/3)$} & $2$ & $-2$ & $-2$& +\\

\hline

{$D_i^c$} & {$({\bf \bar 3, 1},1/3)$} & $-2$ & $-2$ & $-3$& +\\

\hline

{$\mathcal{N}_{2,3}$} & {$({\bf 1, 1},0)$} & $0$ & $4$ & $5$ & +\\

\hline

{$\mathcal{N}_1$} & {$({\bf 1, 1},0)$} & $0$ & $4$ & $5$ & -\\
\hline
\hline
\end{tabular}
}

\caption{Matter fields content of the model, and their charges under the SM gauge group \( G_\text{SM} \), their charges under the local symmetries \( U(1)_{\chi} \), \( U(1)_{\psi} \), \( U(1)_{\psi'} \) and $Z_2$. Here, $Q_{\psi'}=1/4(Q_{\chi}+\sqrt{15} \,Q_{\psi})$, where $Q_{\chi}$ and $Q_{\psi}$ are the normalized GUT generators of $U(1)_\chi$ and $U(1)_\psi$ respectively. Family indices are denoted by the subscript $i\;(=1,2,3)$.}
\label{tab:fields}
\end{table}

Next, we briefly outline how the $\psi'$SM model can arise from an $E_6$ framework. Specifically, the $\psi'$SM can be constructed by following the symmetry-breaking chain shown below (see Ref.~\cite{Lazarides:2019xai} for related discussions),
\begin{equation}
\begin{aligned}
    E_6 &\longrightarrow SO(10) \times U(1)_\psi\\
    &\longrightarrow SU(5) \times U(1)_{\chi} \times U(1)_{\psi}\\
    &\longrightarrow G_\text{SM} \times U(1)_{\chi} \times U(1)_{\psi}\\
    &\longrightarrow G_\text{SM} \times U(1)_{\psi'}\\
    &\longrightarrow G_\text{SM}.
\end{aligned}
\label{seqe6su5}
\end{equation}
The first breaking in Eq.~(\ref{seqe6su5}), following Ref.~\cite{Lazarides:2019xai}, produces an $E_6$ monopole. The subsequent breaking to $SU(5) \times U(1)_{\chi} \times U(1)_{\psi}$ produces an $SO(10)$ monopole charged under $U(1)_{\chi}$ and $SU(5)$ \cite{Lazarides:2023iim}. A third stage of breaking leads to the emergence of the gauge symmetry $G_\text{SM}\times U(1)_\chi\times U(1)_\psi$, along with the formation of a GUT monopole. At this point, a period of cosmic inflation is assumed to dilute or eliminate these monopoles, two of which are, in principle, confined and contribute to the formation of metastable string networks. The fourth stage of symmetry breaking is triggered by a complex Higgs scalar with the same quantum numbers as the RHNs $N_{i}^c$, which breaks $U(1)_\chi \times U(1)_\psi$ down to $U(1)_{\psi^{'}}$.
The residual gauge symmetry $U(1)_{\psi'}$ arises from a linear combination of the $U(1)_\chi$ and $U(1)_\psi$ charges:
\begin{align}
Q_{\psi'}=\dfrac{1}{4(Q_{\chi}+\sqrt{15} \,Q_{\psi})},
\end{align}
which remains unbroken after this fourth stage of symmetry breaking. This specific combination ensures that the $SO(10)$ singlet fermions $\mathcal{N}_i$ carry charges under $U(1)_{\psi'}$, while the RHNs $N_i^c$ remain singlets, as required for the DM and leptogenesis mechanisms discussed below. This step also gives rise to the first class of metastable current-carrying strings.  The electrically neutral current associated with these strings arises from RHN zero modes localized along the string. Here we impose an extra $Z_2$ symmetry under which the lightest $SO(10)$ singlet fermion $\mathcal{N}_1$ is non-trivially charged. This $Z_2$ prohibits the $\mathcal{N}_1$ from decaying, making it a viable DM candidate in our framework. As discussed in Sec.~\ref {sec:DM}, to prevent the DM from thermalization, the relevant couplings must remain sufficiently small. Consequently, the resulting CSs deviate from the conventional Abelian-Higgs type, and current analytical and numerical techniques are not yet adequate to fully capture their dynamical evolution. With the standard current carrying CS formation at this stage, we refer the reader to~\cite{Afzal:2023kqs}\footnote{The spontaneous breaking of $E_6$ to the SM via its maximal subgroup $SO(10)\times U(1)_\psi$ explaining dumbbells, metastable strings, as well as domain walls bounded by necklaces is discussed in Ref.~\cite{Maji:2024pll}.}.
The final breaking, $U(1)_{\psi^{'}}$ is accomplished by a Higgs scalar $\mathcal{N}$ that arises from the $SO(10)$ singlet component of the $E_6$ fundamental representation $27$, sharing the same quantum numbers as the fermionic singlets $N_i$. This breaking leads to a second family of metastable strings, which are superconducting due to the presence of zero modes corresponding to electrically charged exotic matter fields \cite{WITTEN1985557}.

\section{Dark Matter Phenomenology}\label{sec:DM}

In this section, we explore the interesting possibility of having a freeze-in type DM in the present setup. The initial abundance of such DM candidates is assumed to be zero and, as the Universe cools, the DM is expected to be dominantly produced by the decay or scattering of particles present in the thermal bath. By virtue of the tiny strengths of the couplings at play here, the interaction rate(s) is always smaller than the Hubble expansion rate ($\Gamma < H$ , where $\Gamma$ and $H$ respectively denote the relevant decay rate and the Hubble parameter). In the present setup, this possibility is realized if one considers the lightest $SO(10)$ singlet fermion $\mathcal{N}_1$ to be odd under an additional discrete $Z_2$ symmetry. The $Z_2$ symmetry of $\mathcal{N}_1$, ensures its stability over the cosmological time scale. For simplicity, in the rest of the discussion we will use the notation $\mathcal{N}_{\rm DM}$ instead of  $\mathcal{N}_{1}$.

The relevant Lagrangian for the DM phenomenology in the present framework has the form,

\bea
\mathcal{L}_{\rm DM}&=&|D_\mu \mathcal{N}|^2 +\mathcal{N}_{\rm DM}i\gamma_\mu D^\mu \mathcal{N}_{\rm DM}+\frac{c}{\Lambda}\mathcal{N}_{\rm DM}\mathcal{N}_{\rm DM}\mathcal{N}^{\dagger2}+h.c.,
\label{Lag_DM}
\eea
where $D_\mu=\partial_\mu+i g_{\psi}'2\sqrt{10}Q_{\psi'}Z_{\psi'_\mu}$, $c$ is a dimensionless coefficient and $\Lambda$ denotes the effective scale of the theory. The $U(1)_{\psi'}$ symmetry breaking is triggered when the scalar field, $\mathcal{N}$ obtains a non-zero vacuum expectation value (vev), $\langle \mathcal{N}\rangle =v_{\psi'}$.
Consequently, the gauge bosons, $Z_{\psi'}$ and the DM $\mathcal{N}_{\rm DM}$ acquire their respective masses
\begin{align}
M_{\rm DM}&= c\frac{v_{\psi'}^2}{2\Lambda},
\label{DM_mass}
\\
M_{Z_{\psi'}}&= 2\sqrt{10}Q_{\psi'} g_{\psi'} v_{\psi'}.
\label{Zprime_mass}
\end{align}
e
Next, we shall consider the breaking of $U(1)_{\psi^{'}}$ takes place prior to the reheating temperature $T_{\rm RH}$ of the Universe, i.e. during the reheating phase, with $v_\psi' > T_{\rm RH}$ throughout our analysis.  In the context of $E_6$ models, the gauge couplings are typically of order one $(\mathcal{O}(1))$. Consequently, setting $g_{\psi'}\sim \mathcal{O}(1)$ implies that the mass of the associated gauge boson is $M_{Z_{\psi'}}\sim \mathcal{O}(v_{\psi'})$.  As a result, $Z_{\psi'}$ remains decoupled from the thermal plasma for temperatures below $T_{\rm RH}$ given this hierarchy between $v_\psi'$ and $T_{\rm RH}$. This also means that the production of DM from $Z_{\psi'}$ decay can be safely neglected \footnote{If the hierarchy is reversed $i.e~ v_\psi' < T_{\rm RH}$, with $g_{\psi}'\sim\mathcal{O}(1)$, the $Z_{\psi'}$ would be efficiently produced in the thermal plasma and could potentially bring DM into equilibrium. We avoid this scenario, as it would thermalize the DM candidate.  }, as any DM produced in this way would be significantly diluted by the entropy generated during reheating. 

 Once the Universe is reheated, the DM can again be produced as a result of its feeble interaction with the scalar $\mathcal{N}$ if the scalar remains in the thermal plasma. This situation can be realized by appropriately choosing the self-quartic coupling of $\mathcal{N}$. Additionally, a small quartic coupling $\lambda_{\mathcal{N}H}$ between $\mathcal{N}$ and the SM Higgs, together with the large hierarchy $v_{\psi'}\gg v_H$, ensures that the mixing angle $\theta_{\mathcal{N}H} \sim \lambda_{\mathcal{N}H} v_H v_{\psi'}/(M_{\mathcal{N}}^2 - M^2_H)$ remains sufficiently small. This also prevent large quadratic corrections to the Higgs mass, avoids observable deviations in Higgs phenomenology and allows us to neglect $\mathcal{N}$-Higgs mixing in our analysis. Being a freeze-in type DM, $\mathcal{N}_{\rm DM}$ can now be produced through the decay\footnote{The present freeze-in scenario qualitatively resembles the one elaborated in \cite{Biswas:2016bfo} where it is shown that the scattering processes contribute negligibly towards the DM production. The scattering contribution is therefore omitted throughout in the present study.}: $\mathcal{N}\to\mathcal{N}_{\rm DM}\mathcal{N}_{\rm DM}$, if kinematically allowed.
The scalar $\mathcal{N}$ have the following decay width to two $\mathcal{N}_{\rm DM}$ final state:
\begin{align}
\Gamma_{\mathcal{N} \longrightarrow \mathcal{N}_{\rm DM} \mathcal{N}_{\rm DM}}& = \frac{M_{\mathcal{N}}}{16 \pi} \bigg{(}\frac{cv_{\psi'}}{2\Lambda}\bigg{)}^2
 ~\Big(1 - \frac{4 M^2_{{\rm DM}}}{M^2_\mathcal{N}}\Big)^{3/2}.
\end{align}

\begin{figure}[h!]
    \centering
    \includegraphics[width=7.5cm]{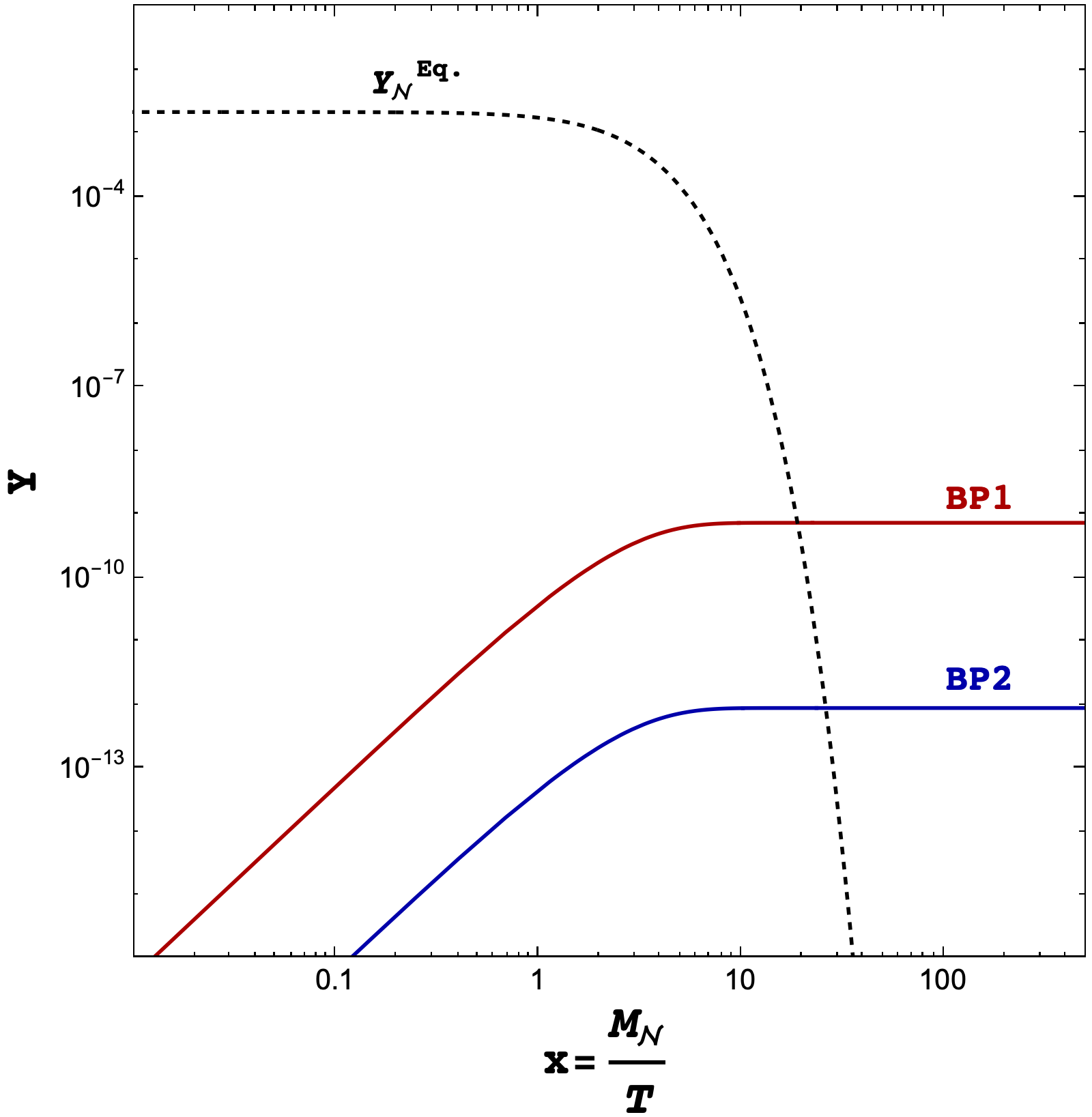}~~~~~~~~~~
    \includegraphics[width=7.7cm]{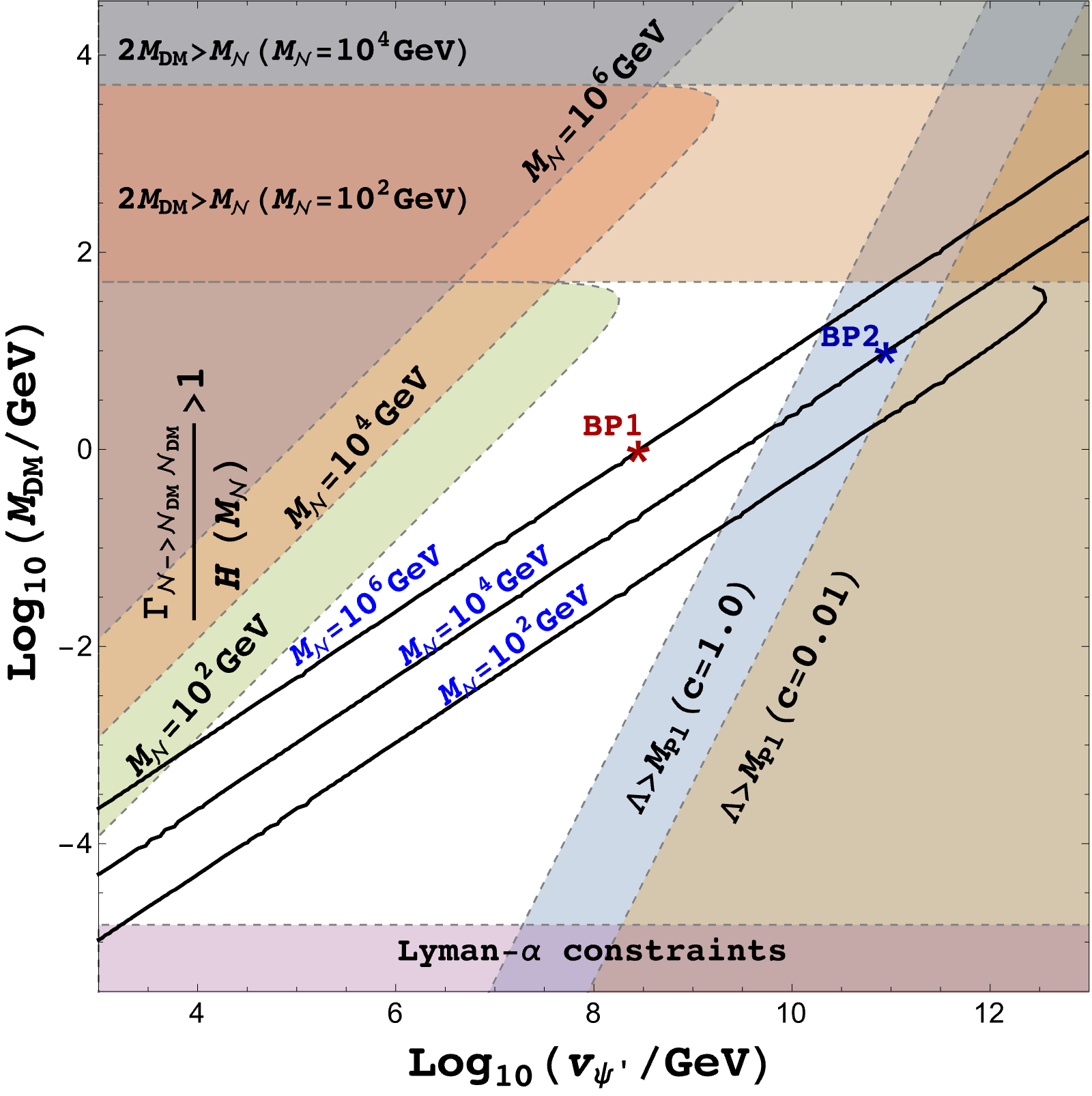}
    \caption{Left panel: Solutions to Eq.~\eqref{BE_FIMP} 
for evolution of $Y_{\mathcal{N}_{\rm DM}}$ for two benchmark points discussed in Table~\ref{tab2}. Right Panel: Relic density allowed parameter space in $M_{\rm DM}-v_{\psi'}$ bi-dimensional plane for $M_{\mathcal{N}} = 10^2~\rm{ GeV}$, $10^4~\rm{ GeV}$ and $10^6~\rm{ GeV}$. The colored region on the left shows the parameter space where the DM can be thermalized, the colored region on the right corresponds to $\Lambda>M_{\rm Pl}$, While the colored region in the top is disallowed from the kinematic condition $M_\mathcal{N}<2 M_{\rm DM}$, the one at the bottom is excluded from the Lyman-$\alpha$ bound.}
    \label{fig:DM}
\end{figure}

\noindent While DM inevitably thermalizes when its couplings to the scalar $\mathcal{N}$ is large, thermalization can be avoided by keeping these couplings sufficiently small,  $i.e ~c v_{\psi'}/\Lambda \ll 1$, hence, we adhere to the condition
$c v_{\psi'}/\Lambda \ll 1$ throughout. This condition helps maintain $\frac{\Gamma_{\mathcal{N}\to \mathcal{N}_{\rm DM} \mathcal{N}_{\rm DM}}}{H(T=M_\mathcal{N})}\ll1$. To study the production and evolution of the DM with the expansion of the Universe, we solve the following Boltzmann equation,

\bea
\frac{d Y_{\mathcal{N}_{\rm DM}}}{d x} &=& \frac{2 M_{\rm Pl}}{1.66 M_{\mathcal{N}}^2} \frac{x \sqrt{g_*(x)}}{g_s(x)} 
   \Gamma_{\mathcal{N} \to \mathcal{N}_{\rm DM} \mathcal{N}_{\rm DM}}Y_\mathcal{\mathcal{N}}^{\rm eq},
 \label{BE_FIMP}
\eea
\noindent where $x=\frac{M_{\mathcal{N}}}{T}$ and $T$ is the temperature of the Universe. While $Y_{\mathcal{N}_{\rm DM}}$ denotes the comoving number density of the DM, $Y_\mathcal{N}^{\rm eq}$ denotes the comoving equilibrium number density of the decaying particle $\mathcal{N}$. The solution of Eq.~\eqref{BE_FIMP} is obtained by setting DM's initial abundance to zero.  In the left panel of Fig.~\ref{fig:DM}, we present the evolution of the DM's comoving number density obtained after solving Eq.~\eqref{BE_FIMP} for two different benchmark points (BP) provided in Table~\ref{tab2}.
\begin{table}[h]
\centering
		\begin{tabular}{|c|c|c|c|c|c|c|c|c|}
			\hline
			& \textbf{$v_{\psi^\prime}$ [GeV]} & \textbf{$M_{\mathcal{N}}$[GeV]} & \textbf{$M_{\rm DM}$[GeV]} & \textbf{$\Omega_{\rm DM} h^2$}   \\
                \hline
            \textbf{BP1} & $ 2.9\times 10^{8}$ & $10^6$ & 1 & 0.12   \\
            \hline
            \textbf{BP2} & $9.2\times10^{9}$ & $10^4$ & 10 & 0.12  \\
            \hline
    		\end{tabular}
		\caption{Two characteristic benchmark values of $v_{\psi'}$, and $M_{\mathcal{N}}$ are listed along with corresponding $M_{\rm DM}$ values that satisfies observed DM abundance. }\label{tab2}
	\end{table}
\noindent As expected, the DM abundance (resulting from the scalar decay) increases slowly with the expansion of the Universe. The abundance of DM freezes-in (saturates) when the equilibrium number density of $\mathcal{N}'s$  becomes Boltzmann suppressed. Once the DM asymptotic abundance $Y_{\mathcal{N}_{\rm DM}}(x=\infty)$ is obtained, one can use it to calculate the DM relic density,

\bea
\Omega_{\rm DM}h^2&=& 2.75\times10^8\bigg{(}\frac{M_{\rm DM}}{\rm GeV} \bigg{)}Y_{\mathcal{N}_{\rm DM}}(x=\infty).
\eea

\noindent The BPs presented in Table~\ref{tab2} satisfies the observed DM abundance $\Omega_{\rm DM}h^2=0.12\pm 0.0012$ \cite{Planck:2018vyg}. Finally, in the right panel of Fig.~\ref{fig:DM}, we show the allowed parameter space that remains consistent with the observed DM relic density shown by the black solid lines for three different masses of $\mathcal{N},~i.e.~ M_{\mathcal{N}}=10^2~\rm{GeV}, 10^4~\rm{GeV}$ and $10^6~\rm{GeV}$ respectively in the $M_{\rm DM}-v_{\psi'}$ plane. The two BPs are shown by the red and blue stars.  While the coloured region on the left show the parameter space where the DM can be thermalized, the coloured region on the right shows $\Lambda>M_{\rm Pl}$ for two different values of $c=1.0$ and $0.01$. While the colored region in the top is disallowed from the condition $M_\mathcal{N}<2 M_{\rm DM}$, the one at the bottom is excluded from the Lyman-$\alpha$ bound \cite{Decant:2021mhj}.  As evident from the plot, the current setup allows for a DM candidate spanning a wide mass range, provided it is produced from the decay of $\mathcal{N}$.  Although, we do not show a plot in $M_{\rm DM}-T_{\rm RH}$ plane, it can easily be obtained by replacing $v_{\psi'}$ with $T_{\rm RH}$ in the right panel of Fig. \ref{fig:DM} by choosing a certain (mild) hierarchies between $v_{\psi'}$ and $T_{\rm RH}$.

\section{Neutrino masses and Leptogenesis}\label{sec:NM_lepto}

\begin{figure}[htb!]
    \centering
    \includegraphics[width=13cm]{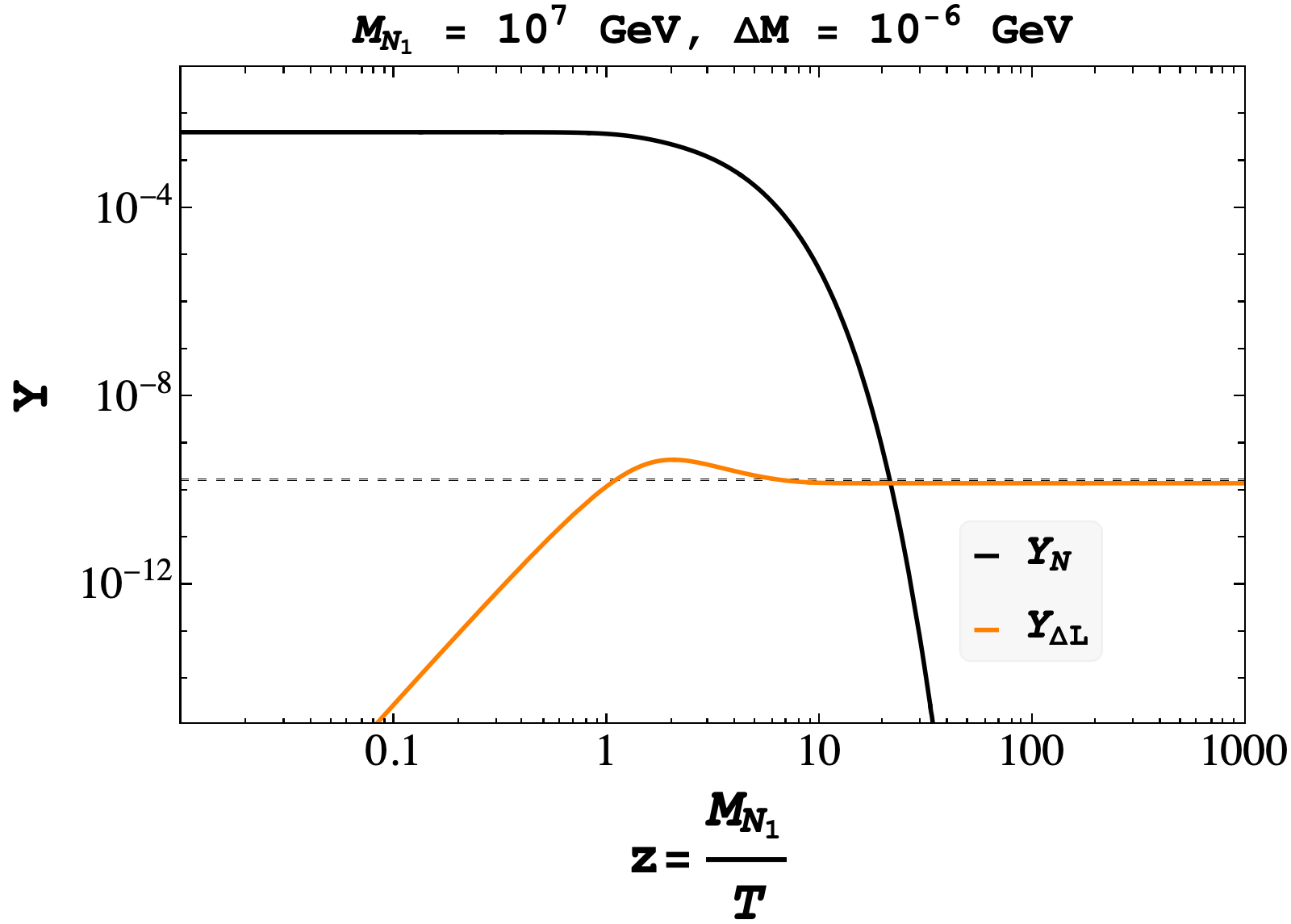}
    \caption{Solutions to BEQs (Eq. \eqref{eq:be_asymm}) 
for evolution of $Y_{N_1}$
(black), and lepton asymmetry $Y_{\Delta L}$
(orange). The gray dashed line indicates the correct
 $Y_{\Delta L}$ is required to produce the observed baryon asymmetry of
the Universe. Here we have set $z_R=0.05 -0.9i$.}
    \label{fig:baryon_asy}
\end{figure}

 Although not connected to the DM, the present setup also provides a possibility of neutrino mass generation via the type-I seesaw mechanism, together with a possible explanation of the baryon asymmetry of the Universe from the RHN's decay. Importantly, the flexibility in the reheating temperature arising from our DM considerations opens up a wider range of possibilities for baryogenesis scenarios involving RHNs. As inferred from the DM phenomenology of the previous section that the reheating temperature in our
case can fall in a broad range. In connection to this, it is interesting to recall that the standard thermal leptogenesis works for the RHN mass above $10^9$ GeV, the so-called Davidson-Ibarra bound \cite{Davidson:2002qv}. However, this conclusion is based on the hierarchical nature of RHNs and often requires a minimum $T_{\rm RH}>10^9$ GeV.
If the RHNs are nearly degenerate, as in resonant leptogenesis \cite{Pilaftsis:1997jf,Pilaftsis:2003gt}, the asymmetry production can be significantly enhanced. In this scenario, both the RHN mass scale and the required reheating temperature can be significantly reduced, thereby enlarging the viable parameter space for successful leptogenesis. Therefore, in this work, we focus on resonant leptogenesis as the mechanism for explaining the baryon asymmetry of the Universe.  

For the purpose of analysis, we consider two of the RHNs to be present in the thermal bath requiring $T_{\rm RH} > M_{1,2}$ and a sizeable neutrino Yukawa interactions so as they can be responsible for thermal leptogenesis, while the $N_3$ to be superheavy so that it is decoupled from the rest. Assuming the $N_1$ and $N_2$ to be nearly degenerate, we define their mass splitting as $\Delta M=M_2-M_1$, with $\Delta M/M_i<<1$. The CP asymmetry parameter $\epsilon_{N_i}$ associated to the decay of the $i$-th RHN can be expressed as 
	\beeq
	\epsilon_{N_i} = \frac{\sum_{\alpha}[\Gamma_{N_i\to L_{\alpha} H}-\Gamma_{N_i\to \bar{L}_{\alpha} \bar{H}}]}{\sum_{\alpha}[\Gamma_{N_i\to L_{\alpha} H}+\Gamma_{N_i\to \bar{L}_{\alpha} \bar{H}}]},
	\eeq
	which results from the interference of the tree-level decay of $N_i$ and the one loop, vertex and self energy, diagrams. The general expression for such CP asymmetry {(after flavor sum)} can be estimated as \cite{Flanz:1996fb,Pilaftsis:1997jf,Pilaftsis:2003gt,Iso:2010mv,Qi:2022fzs,Das:2024gua}
	\beeq
	\epsilon_{N_i}=-\sum_{j\neq i}\frac{M_i \Gamma_{N_j}}{M_j^2}\left(\frac{V_{ij}}{2}+S_{ij} \right) \frac{{\rm Im}(y_{\nu}^{\dagger}y_\nu)^2_{ij}}{(y_{\nu}{^\dagger}y_\nu)_{ii}(y_{\nu}^{\dagger}y_\nu)_{jj}},
	\label{CP_asy}
	\eeq
	where
	\beeq
	V_{ij}=2\frac{M_j^2}{M_i^2}\left[\left(1+ \frac{M_j^2}{M_i^2}\right) {\rm ln}\left(1+ \frac{M_i^2}{M_j^2} \right)-1  \right],
	\eeq
	\beeq
	{\rm and} ~~~S_{ij}=\frac{M_j^2(M_j^2-M_i^2)}{(M_j^2-M_i^2)^2+M_i^2\Gamma_{N_j}^2},
	\eeq
	denote the vertex correction and self energy corrections respectively. At this stage, it is important to point out that though the contributions of the vertex and self energy corrections are of similar order for hierarchical RHNs, the self energy contributions dominate over the other in case of quasi-degenerate RHNs. This is mainly because, in this limit, the CP asymmetry can be maximised with $M_i^2-M_j^2 \sim M_i \Gamma_{N_j}$ such that $S_{ij} \gg V_{ij} \sim \mathcal{O}(1)$ \cite{Pilaftsis:2003gt}. Next, ${y}_{\nu}$ is the neutrino Yukawa coupling matrix in the mass diagonal basis of RHNs, the form of which can be obtained using Casas Ibarra (CI) parametrization \cite{Casas:2001sr} :
\begin{equation}
{y}_\nu = \frac{\sqrt{2}}{v_h}U_{\rm PMNS}^*\sqrt{m_\nu^d}{\cal R}^T\sqrt{M_R}\, ,
\label{y-nu-CI}
\end{equation} 
where $M_R (m_\nu^d)$ represents the diagonal RHN (light neutrino) mass matrix, while 
$U_{\rm PMNS}$ \cite{ParticleDataGroup:2020ssz} is the unitary matrix (in charged lepton diagonal basis) required to diagonalise $m_\nu = U_{\rm PMNS}^\ast \,m_\nu^d\, U_{\rm PMNS}^\dagger$ and $v_h=246$ GeV. 
Here, ${\cal R}$ is a $3 \times 3$ orthogonal matrix that can be chosen as \cite{Antusch:2011nz} :
\begin{align}
{\cal R} =
\begin{pmatrix}
0 & \cos{z_R} & \sin{z_R}\\
0 & -\sin{z_R} & \cos{z_R}\\
1 & 0 & 0
\end{pmatrix} ;
\end{align} 
where $z_R=a+i\,b$ is a complex angle. It is important to note that the model provides sufficient flexibility in choosing the parameter $ z_R$ to produce correct leptogenesis 
and neutrino mass, given $M_{N_1}$. 

Next, we elaborate on the evolution of number densities via the Boltzmann equations given below,

\begin{eqnarray}
 \frac{dY_{N_i}}{dz}&=&-\frac{1}{sHz}\gamma_{N_i}\bigg(\frac{Y_{N_i}}{Y_{N_i}^{\rm eq}}-1\bigg)\,,\nonumber\\
 \frac{dY_{\Delta L}}{dz}&=&\frac{1}{sHz}\bigg[\gamma_{N_i} \bigg\{\epsilon_i \bigg(\frac{Y_{N_i}}{Y_{N_i}^{\rm eq}}-1\bigg)-\frac{Y_{\Delta L}}{2Y_l^{\rm eq}}\bigg\} \bigg]\, \label{eq:be_asymm},
\end{eqnarray}
where the yield is defined by $Y^{\rm eq}=\frac{n^{\rm eq}}{s}$, ($n^{\rm eq}$ is the (equilibrium) number density, $s=0.44 g^\ast T^3$ is the total entropy density); and $z= {M_{N_1}} /T$, where $T$ is temperature.
The reaction density $\gamma$ is given by :
\bea
\gamma(a \rightarrow bc) = n^{\rm eq}\frac{K_1(z)}{K_2(z)}\Gamma(a \rightarrow bc),
\eea
where $K_{1,2}$ are Bessel functions of 1st, 2nd kind. Assuming the RHNs to be in equilibrium with the thermal plasma in the early Universe, we set the initial condition as $ Y_{N_1}=Y_{N_1}^{\rm eq}, Y_{\Delta L}=0$. 
We would like to point out that the benchmark values of $M_1, \Delta M$ and $z_R$ are chosen in a way to obtain the correct amount of baryon asymmetry via resonant leptogenesis from the subsequent decay of $N_i$ 
with a fixed value of $M_{N_1} =10^{7}$ GeV. The choices of these parameters also ensure correct neutrino mass generation via 
CI parametrization as described above. The numerical solution to BEQs for $M_{N_1} = 10^{7}{\rm~GeV}, \Delta M = 10^{-6}{\rm~GeV}$ and $z_R=0.78+30i$ is shown in the Fig.~\ref{fig:baryon_asy}.  The choice presented above is merely one illustrative
example.

Turning to $Y_{N}$ (Eq.~\ref{eq:be_asymm} and black thick line in Fig. \ref{fig:baryon_asy}), processes that contribute significantly to the RH abundance are  
$N_i \rightarrow l H\,, \quad N_i \rightarrow \bar{l} \bar{H}\,, \quad l H \rightarrow N_i\,, \quad \bar{l}\bar{H}\rightarrow N_i\,$. They keep $Y_{N_i}$ into equilibrium . Decay of $N_i$ to $\ell H(\bar{\ell}\bar{H})$ is responsible for generating lepton asymmetry $Y_{\Delta L}=Y_l-Y_{\bar{l}}$ described via the second equation of 
Eq.~\ref{eq:be_asymm} and shown by the orange thick line in Fig.~\ref{fig:baryon_asy}. $Y_{\Delta L}$ being proportional to $\epsilon_i$ 
(first term in Eq.~{\ref{eq:be_asymm}}), is responsible for the rise in asymmetry, which gradually fades due to washout by inverse decays 
$lH(\bar{l} \bar{H})\rightarrow N_i$ denoted by the second term in Eq. ~\ref{eq:be_asymm}. As temperature falls below $M_1$, the washout 
processes get suppressed and once $N_i$ decays are complete, the asymmetry saturates (grey dashed line). The asymptotic 
yield $Y_{\Delta L}^0$ is eventually transferred to baryons ($Y_B$) (via electroweak sphalerons above $T \sim 100 {\rm~GeV}$) following, 
$Y_B = c Y_{\Delta L}^\infty$, with $c=28/79$ to produce $Y_B=(8.75 \pm 0.23) \times 10^{-11}$. Here we do not incorporate the flavor effects on leptogenesis\footnote{{For flavor effects on thermal leptogenesis, we direct the readers to \cite{Abada:2006fw,Nardi:2006fx,Blanchet:2006be,Dev:2017trv,Datta:2021elq,Datta:2021gyi} and for flavor effects on leptogenesis during a non-instantaneous reheating epoch, we refer the readers to \cite{Datta:2022jic,Datta:2023pav}.}} for simplicity.

\section{Conclusion}\label{sec:conclusion}
We explore the freeze-in production of DM and the generation of the matter–antimatter asymmetry within a $\psi'$SM model framework. This scenario emerges through the symmetry-breaking chain: $E_6\to SO(10)\times U(1)_\psi\to SU(5)\times U(1)_\chi\times U(1)_{\psi}\to G_{\rm SM}\times U(1)_\chi\times U(1)_{\psi}$. The $U(1)_{\psi'}$ symmetry present in the setup results from the linear combination of $U(1)_\chi$ and $U(1)_\psi$ contained in $E_6$. While RHNs present in the 16-plet of $SO(10)$ does not carry any charge under the $U(1)_{\psi'}$, the singlet fermions of $SO(10)$ carry a non-zero charge (with $2\sqrt{10}\,Q_{\psi'}=5$). This charge forbids them from acquiring masses through renormalizable interactions. Their masses are instead generated via a dimension-five operator involving singlet fermions $\mathcal{N}_i$ and scalar $\mathcal{N}$, once $U(1)_{\psi'}$ is spontaneously broken. Further a discrete symmetry is also incorporated to ensure the stability of the lightest fermion $\mathcal{N}_{\rm DM}$. We show that, depending on the size of the breaking scale of $U(1)_{\psi'}$ and the mass of the scalar $M_{\mathcal{N}}$, a DM with mass ranging from a few MeV to a few hundred GeV can easily be produced via freeze-in mechanism from the decay of scalar $\mathcal{N}$. A strict upper bound on the DM mass is obtained by implementing $\Lambda<M_{\rm Pl}$. It is worth noting that while the extremely feeble interactions between this DM candidate and SDM particles make it challenging to probe through direct detection, indirect searches, or collider experiments, our framework offers a promising alternative. Gravitational waves from cosmic strings provide a viable detection pathway that could reveal the $U(1)_{\psi'}$ symmetry breaking scale, thereby constraining the dark matter mass which we aim to explore in our future work. The RHNs in this framework can have Majorana masses, enabling the generation of small neutrino masses through the standard type-I seesaw mechanism and simultaneously facilitating the matter–antimatter asymmetry via resonant leptogenesis. 

\section*{Acknowledgments}
The authors thank Qaisar Shafi and Rinku Maji for the useful discussions and interesting comments on the manuscript. RR acknowledges the STFC Consolidated Grant ST/X000583/1.

\bibliographystyle{apsrev4-1}
\bibliography{refs}
\end{document}